# Slicing functional aspects out of legacy applications

Barthélémy Dagenais[1]
*School of Computer Science*
*McGill University*
*Montréal, QC, Canada*
*bart@cs.mcgill.ca*

Hafedh Mili
*LATECE Laboratory (www.latece.uqam.ca)*
*Université du Québec à Montréal*
*Montréal, QC, Canada*
*hafedh.mili@uqam.ca*

## Abstract

*Aspect-oriented software development builds upon object-oriented development by offering additional abstraction boundaries that help us separate different types of requirements into so-called aspects. Given a legacy OO application, it pays to identify existing aspects to help understand the structure of the application and potentially to extract those aspects into separate software modules that can be maintained and reused individually. We present an approach to extract functional aspects using program slicing. We first characterize what we mean by functional aspect, and explore the extent to which program slicing techniques could improve over existing aspect mining and feature location methods for such aspects. We then describe the results of applying our approach to two medium-sized open source programs. Our preliminary results show that program slicing can yield well-defined functional aspects when the appropriate slicing criteria are used. Finally, we explore the effect of design quality on the candidate functional aspects.*

## 1. Introduction

Software applications typically embody a complex web of requirements, both functional ones related to the <input, output> relation implemented by the software, and non-functional ones, related to how that output is produced. A good software abstraction and packaging technique is one that enables us to implement different requirements in distinct software artefacts that we can develop, maintain, and compose at will [14]. Aspect-oriented software development (AOSD) builds upon the abstraction and packaging techniques inherent in object-oriented software by proposing new software artefacts that enable us to untangle more types of requirements than object-oriented abstractions allowed. We use the term aspectual artefact to mean an aspect *à la* AspectJ™ [13], a hyperslice *à la* HyperJ™ [19], a composition filter [1], or any other artefact proposed by any of the AOSD techniques. Given a legacy object-oriented application, it pays to identify and to try to isolate code fragments that implement a particular requirement: minimally, this can help in understanding the application, and by extension, in maintaining it by delineating the parts of the application that need to be modified or that are affected by a change in requirements. In some cases, it may even be possible to repackage those fragments using object-oriented refactoring [8] or AOSD-like refactoring [15] so they can be reused and composed with other programs. Our work deals with aspect mining in legacy Java code.

It is customary in the AOSD community to make a broad distinction between two families of aspects: one derived from subjects [10] and one derived from AspectJ [13]. Although both families can possibly implement any kind of requirement, AspectJ-like aspects are often used to implement design level requirements such as persistence, distribution, security, logging, whereas subject-like aspects are often used to implement pluggable features that depend on core functionality [10]. A textbook example of subjet-like aspects concerns a telephone switching system: the basic functionality of such a system consists of routing calls and managing connexions, but we can have additional features (subjects) such as call-waiting, three-way calling, call forwarding, etc. We are interested in the identification, packaging, and reuse of functional aspects in the form of subject-like aspects that implement functional requirements. Such aspects embody often undocumented domain knowledge that is buried in legacy application code.

Existing aspect mining methods are typically heuristic in nature [5], and look for static or run-time code patterns that characterize occurrences of AspectJ-

---

like aspect instances. However, the identified code fragments are neither cohesive nor complete. Feature location techniques such as SNIAFL [24] are also based on heuristics, and suffer from similar problems, even though the code fragments tend to be more cohesive. The question now is how to uncover such aspects and improve over those techniques. Because everything that is used within a subject-like aspect is also declared within it, we had the intuition that program slicing techniques could be used to precisely identify those fragments of a legacy application that correspond to a functional aspect. This would provide the robustness (soundness of program slicing over fuzziness of heuristics) and completeness (slice can be executable) that is often required for maintenance and repackaging.

In section 2, we describe more precisely what we mean by functional aspect, and explore, through the analysis of a so-called 'refers-to graph', the challenges in extracting such aspects. In section 3, we discuss how program slicing can be used to extract functional aspects and address the challenges discussed in section 2. Section 4 presents the results of applying our approach to two mid-sized open-source projects. We discuss the impact of design quality on the extraction process in section 5. Related work is presented in section 6, and we conclude in section 7.

## 2. Characterizing functional aspects

We are interested in functional aspects that look like subjects [10] or their descendents, hyperslices [19], for two main reasons. First, historically, subject-oriented programming [10] and its descendant [19] have been motivated by the need to separate functional aspects or roles that are played by the same entities, whereas aspect-oriented programming has been motivated by the need to untangle architectural or design-level aspects. Thus, it is reasonable to expect that functional aspects will look more like subjects than like aspects. Second, subjects are partial object-oriented applications, and as such, are meant to be independently compilable—if not executable—which enables us to considerably reduce the search space. In this section, we will look at how such functional aspects will manifest themselves in legacy code developed without AOSD techniques.

Consider the example of the class **Truck** and the various roles it plays within a merchandising company that uses trucks to deliver customer orders. Different departments within the company (accounting, operations and maintenance) will have different views on the functionality of the Truck class. Figure 1 shows the three perspectives represented in their own class hierarchies.

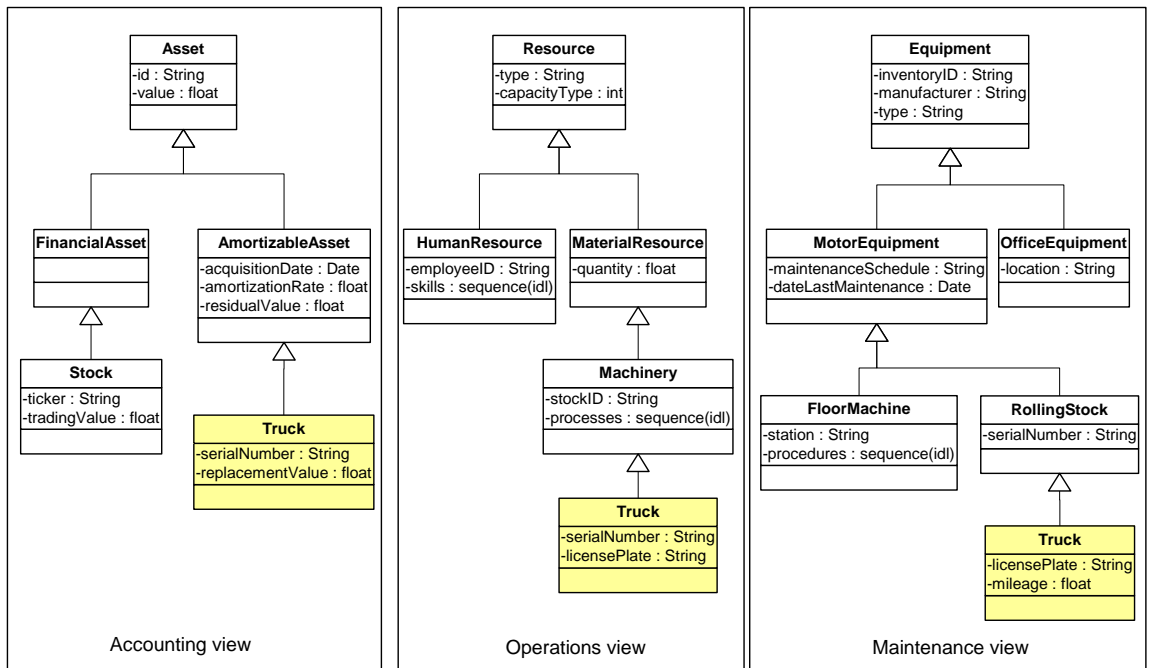

**Figure 1. The Truck hierarchy and its three roles.**

In the absence of AOSD techniques, we would implement all three sets of functionalities in the same class hierarchy using traditional OO programming and design idioms, including multiple inheritance or composition. Whichever solution was used to implement the class Truck, the question is then to recognize within the legacy class hierarchy the three functional roles that it embodies. More specifically, given a legacy class hierarchy that presumably embodies several functional aspects, how can we isolate the set of classes, attributes, and functions that, together, implement one such functional aspect.

As this example illustrates, some attributes will be common to all three functionalities (e.g. 'serialNumber') while others will be *exclusive* to a particular view (e.g. 'amortizationRate' used in accounting view), or a subset of views (e.g. 'licensePlate' used in operations and maintenance views). The same is true for functions (not shown in Figure 1). The functions that are *exclusive* to a particular functional aspect will only refer to attributes and other functions that are *accessible* to that aspect, meaning, those that are *exclusive* to it, as well as those that are shared between this aspect and others.

By contrast, the functions that are shared between several functional aspects may refer to attributes or functions that are not visible to some of them. A method such as 'toString()' or 'save()' would typically be shared by all aspects and refer to all attributes, including ones that are exclusive to some aspects.

In the general case, functional aspects may share attributes and functions. We see two types of function sharing, 1) sharing semantics, and 2) sharing semantics and code. In the first case, the function would have the same signature but different aspects (and combinations thereof) will have different bodies. In the second case, signature and body are common. An example of the former is the 'toString()' method which, for each aspect, concatenates the values of the attributes that are accessible (exclusive or shared) to that aspect. An example of the latter is an accessor to a shared attribute (e.g. 'getSerialNumber()'). Functions of the first kind are called *multiply-defined functions* whereas functions of the same kind are called, simply, *shared functions*.

These functions need to be handled differently when we try to extract functional aspects. A shared function will be made part of every functional aspect, as is. A multiply-defined method may need to be sliced to recover those parts that are specific to a functional aspect. Figure 2 illustrates the general case. The three aspects share three attributes and one function (with vertical stripes). They also have a multiply-defined method (oblique stripes) that refers to one shared attribute, in addition to referring to aspect-specific attributes.

The "refers-to" relation defines a binary relationship between the set of functions—call it F—and the union of the set of attributes—call it A—with the set of functions F. In other words:

$$\text{refers-to}: F \rightarrow A \cup F$$

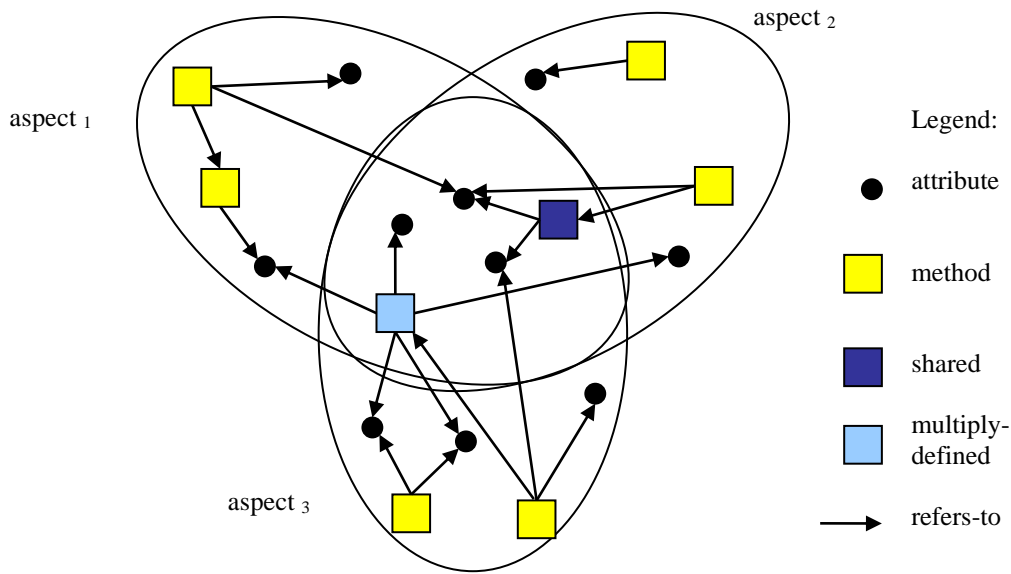

**Figure 2 Three functional aspects**

The question then becomes: given the shape of the refers-to relationship graph that results from a legacy class hierarchy that is known to implement several functional aspects, can we recognize subsets of functions and attributes that correspond to individual functional aspects?

Our first intuition was that the transitive closure of the "refers-to" relationship will yield cohesive subsets of functionality, and that a functional aspect will include one or several such subsets. However, a closer look at the "refers-to" relationship will reveal that:

1. Notwithstanding multiply-defined methods, if we start with a function that is exclusive to a functional aspect, the slice generated by the transitive closure of the refers-to relationship will gather what that function needs, but there is no general relationship between the slices generated by functions belonging to the same functional aspect, versus slices generated by functions belonging to different aspects. Indeed, this is illustrated by aspect2 in Figure 2 which is made of two disjoint subcomponents: in this case, if we start from one component, there is no way with the refers-to relationship to uncover the second subcomponent.
2. With multiply-defined methods, all bets are off: a slice can start with a function that is exclusive to a functional aspect and stay within that aspect until it hits a multiply-defined method (e.g. 'save'), in which case it veers off to functions and attributes belonging to other aspects.

Thankfully, program slicing techniques can help in both situations, as we show in the next section.

## 3. Program slicing

### 3.1. Slicing flavors

Program slicing is a code analysis technique developed by Mark Weiser in his PhD thesis [22]. The original definition of a program slice with respect to a program point p and a variable v consists of the set of statements of the program that can potentially affect or be affected by the value of v at point p [11]. The pair <p, x> is called slicing criterion. Program slicing has many applications, including helping in code understanding, testing, debugging, or in change impact analysis [11].

It is customary in the literature to distinguish between static and dynamic slicing. Static slicing uses only information that is available at compile time whereas dynamic slicing computes a slice based on a particular program execution, typically reducing the size of the slice significantly. Another common distinction concerns the direction of the dependency relationship: backward slicing looks for statements that influence a variable v at a program point p, whereas forward slicing determines the set of program statements that are influenced by the value of the variable v at program point p.

The kind of dependencies between program statements, and the efforts required to uncover them depend heavily on the underlying programming language. Languages that use dynamic memory allocation and pointers also raise a number of challenges. Object-oriented programming languages came with their own set of challenges, including polymorphism and dynamic binding. One solution to this problem consists of performing pointer analysis as in [16] in order to reduce the set of possibilities. Other features such as reflection, concurrency [17] and exception handling [2] raise their own sets of issues. Since we want to extract all code related to the output of a function, we are more interested in static backward slicing.

### 3.2. Using slicing to extract functional aspects

In section 2, we attempted to characterize a functional aspect of a class hierarchy as the subset of methods and attributes that together, support a coherent set of functions. For that characterization, we relied on dependencies between functions and attributes (a function refers to a function or to an attribute). Consider now what a function from the accounting view of Figure 1 such as 'computeResidualValue()' might look like (see Figure 3). In this case, our starting point is a function that is known to belong to a particular functional aspect, and our objective is to find out which other class attributes or functions it requires. Thanks to the references to the data members 'purchaseDate', 'purchaseValue', and 'amortizationRate' inside the function body, we are able to infer that the accounting functional aspect includes the function 'computeResidualValue()' and those data members. With slicing, the starting point is a variable at a particular program point, and the outcome is the set of program instructions that influence that variable. Thus, if we wanted to find out which other function and data members the function 'computeResidualValue()' requires, we can slice the program based on the results produced by the function. Then, from the set of program statements extracted by the slicing program, we record the data and function member declarations and add them to the functional aspect extension.

However, there are two problematic situations. First, if the function at hand does not return a value, what slicing criterion should we use to identify the data and function members that it uses? For example, instead of returning a value, the function 'computeResidualValue()' might call a setter with the result as illustrated by the commented statement. In this case, we will miss out on the data member 'residualValue'. The second situation is illustrated by the call to printObject(), and works to the advantage of program slicing as a way of extracting functional aspects. In this case, a simple 'refers-to' analysis of the 'computeResidualValue()' function will collect all the functions and attributes it refers to, including ones that are not relevant to the computation of the residual value. In this case, it would collect the 'printObject()' function, and all of the attributes 'printObject()' refers to—i.e. likely, *all* the attributes of the class. This is due to the lack of functional cohesion of 'computeResidualValue()', which does more than updating the residual value, it displays the object. Thankfully, backward slicing on the residualValue attribute will identify the proper subset of attributes and function members.

The example of printObject() also illustrates the problematic case of what we called multiply-defined functions, which are functions that are called from various functional aspects, and that access attributes from all aspects. Those functions will wreak havoc on a 'refers-to' (or def-use) analysis, but will not disrupt slicing as much, provided that we select the right slicing criterion. In fact, those functions will, themselves, get sliced.

In summary, we hypothesize that:
1. program slicing may be used to extract functional aspects, one function at a time
2. the quality of the slice will depend on the selection of the slicing criterion (a disadvantage)
3. program slicing will weed out references to data and function members that are not relevant to the computation of a particular result (an advantage).

In the next section, we describe the tool set that we used for our case studies.

```
class Truck {
…
private float purchaseValue;
private Date purchaseDate;
private float amortizationRate;
private float residualValue;
private Date nextMaintenance;
…

// update residual value and print.
public void computeResidualValue(){
   float resValue = 0;
   int nbrYears = yearsSince(
   purchaseDate);
   resValue = purchaseValue *
      Math.exp(1-amortizationRate,
         nbrYears);
   printObject();
   residualValue = resValue;
   // setResidualValue(resValue);
}
…
public void printObject() {
   display(purchaseValue);
   …
   display(nextMaintenance);
}
…
}
```

**Figure 3. Slicing the accounting functional aspect**

### 3.3. The UQAM Toolset

A lot of work has been done on program slicing over the past two decades, but there are only a few implementations that are publicly available. For the purposes of our research, we used Indus, a complete modular Java static slicer [12].

Indus uses a slightly different approach from the traditional System Dependence Graph [11]. It considers slicing as one of many program analysis applications that rely on analyzing and marking statements in the code. As such, it may be seen as a code analysis framework, with slicing being one instantiation of the framework. It performs different static analyses that mark statements in the code. It is up to the user to decide what to do with the selected statements: they may decide to generate xml slices, executable class slices or other kind of slices from the marked statements.

Indus analyses are performed on Jimple, an intermediate representation of Java programs based on the SOOT framework [21]. Roughly speaking, each bytecode is mapped to a Jimple instruction that uses typed local variables, in lieu of stack addresses. One of the good side effects of using an intermediate representation is that it enables us to handle functional aspects coupled at the statement level in a multiply-defined method. Indeed, even if multiple variables are used within a java statement, they will be decomposed into multiple Jimple statements each of which possibly belonging to a different functional aspect.

For the purposes of understanding and reusing functional aspects, it was clear that the output of Indus—Jimple or bytecode slices—was not adequate. Thus, we extended Indus to generate the corresponding Java source code from the Jimple slices. This was a non-trivial task for several reasons, mainly because the mapping from Java to bytecode, and by extension to Jimple, is one to many. Our extension used the Eclipse Java Development Tool [6] to create the sliced Java source code based on the original source code and the results from the slicer.

In addition to the Java code generator, we built a slice diff and reporting tool that can extract the code, fields, methods and classes that are common to a set of slices or that are unique to a particular slice. From these, the tool generates a dynamic HTML report that presents the different views of a class.

## 4. Case study

In section 3.2, we hypothesized that program slicing may be used to extract functional aspects, one function at a time, provided that we select the appropriate slicing criterion. We also argued that program slicing should yield functional aspects that are more precise than the ones generated by refers-to analysis. The purpose of these case studies is to validate the first hypothesis by applying our technique to mid-sized applications that are complex enough to cover a wide variety of situations and to show the scalability of the approach, but that are also small enough and reasonably well-documented to allow us to inspect the generated slices manually, and assess their quality.

The first challenge is to find good test data. Quantity is not an issue: sourceforge.net alone hosts over 147,000 open source projects. Our search criteria included: 1) the programming language (Java), 2) the size (between 10,000 and 100,000 lines of code), 3) quality (of source code and documentation), 4) the likelihood of supporting several functional domains or aspects, and 5) tractability by program slicing techniques. On the likelihood of supporting several functional aspects, we looked for applications with some complexity, possibly with different invocation modalities, and with average functional cohesion. Loosely coupled application frameworks or class libraries would probably be too functionally sparse to be instructive. Similarly, single monolithic applications would probably be too cohesive to produce distinct functional aspects. The issue of tractability added more constraints. First, because program slicing relies on control dependencies, event-driven systems are known to derail slicing algorithms[2], and so we should avoid projects that can only be invoked through GUI components. Second, program size and the availability of source code preclude the use of projects that use large proprietary libraries[3].

We found two such projects: the JReversePro, and ProGuard. Both are Java (byte)code manipulation utilities that offer a command-line interface. The JReversePro case study is described in section 4.1, ProGuard in section 4.2, and lessons learned from both studies in section 4.3.

### 4.1. The JReversePro application

As stated on the JReversePro website: "JReversePro is a Java Decompiler / Disassembler written entirely in Java (…) The ultimate objective of this project is to provide a decompiler that generates a Java object-based structure that can be programmatically inspected using a specific API." JReversePro is relatively small (85 classes and 12 000 LOC) and does not depend on any external proprietary library: all the functionalities were implemented in the code. Finally, the main functionalities are well defined, making it possible to manually (visually) inspect a slice and assess its quality.

JReversePro provides three functions, constants pool retrieval, class disassembly, and class decompilation. The constants pool retrieval function is simple, and we expected a slice for constants pool retrieval to comprise only a small part of the entire application. By contrast, class disassembly is by definition more involved as it requires a more elaborate analysis of class files, and a parsing of the bytecode. Thus, we expected the disassembly slice to be bigger than the first. Finally, class decompilation requires not only class file parsing (as for class disassembly) but also Java code generation, and we expected it to be yet bigger than the other two.

The first issue we faced was to select the slicing criterion to use to extract each functional aspect. As it turns out, JReversePro has a single main class (JCmdMain) to handle all three functions, and those functions are dispatched in a process method with a conditional structure based on the operation code given on the command line:

```
if (OP_CODE == DISASSEMBLE){
  result = <CODE TO DISASSEMBLE>
} else if (OP_CODE == DECOMPILE) {
  result = <CODE TO DECOMPILE>
} else if (OP_CODE == CONSTANTS_POOL) {
  result = <CODE FOR CONSTANTS POOL>
}
  print result;
```

---

[2] Notwithstanding theoretical difficulties due to what Ranganath et al. called *non-termination sensitive control dependencies* [18], there are practical ones: even if we are able to tie event dispatchers to event handlers, the number of event handler implementations, for a given event listener interface will send the slicer off the deep end.

[3] Indus has a way of dead ending analyses when we get into some code base. This is not a major issue with libraries, but becomes one with *frameworks*, because their components can call the user's code (the Hollywood principle).

For the constants pool retrieval functionality, we selected the line that contains the invocation of the constant pool retrieval function as the slicing criterion (boxed in the above figure). The produced slice seemed accurate. Only 20 classes (out of 85) were part of the slice and by looking at the remaining classes, methods, fields and code, it was clear that only the constants pool retrieval functionality was retrieved.

We used a similar approach for the disassembling and decompiling functions. After slicing the two functionalities, we found that the two slices were identical. This did not seem correct, because we knew that decompiling involved a lot more work than disassembling. We looked at the slicing criteria and we found that the function calls for the two functionalities were similar, except for one parameter value: a control flag. We followed the flag in the call hierarchy and found that the slicer was fooled again by a conditional statement:

```
JreverserRengineer jre;

if (flag){
  jre = new JDisAssembler();
} else {
  jre = new JDecompiler();
}

jre.execute(…)
```

The slicer we used is able to correctly analyze pointers so it would not have been fooled by a polymorphic assignment such as

jre = new JDisAssembler();

and it would have been able to gather only the relevant classes and methods. The problem here was the conditional statements.

To circumvent the problem, we had to manually change the initialization of JDecompiler to a null value when we sliced the disassembling function and vice versa for the slicing of the decompiling function. By doing so, the slicer yielded satisfactory results: the decompiling functionality included almost all the code found under the disassembling functionality (30 classes). It also included additional functionality (51 classes in all). The classes that were never included were part of the optional GUI interface for JReversePro.

Overall, more than half of the classes (55%) participated in more than one functional aspect, and about 30% of those used different sets of fields and methods for each functional aspect. Further, we noticed that small methods tended to be highly cohesive, and were either included in a (functional) slice in their entirety, or were excluded altogether. By contrast, bigger methods are sometimes sliced, with some fragment included in functional aspect A and another in functional aspect B. These correspond to what we called multiply-defined methods in sections 2 and 3.2.

### 4.2. The ProGuard application

The ProGuard website states that "ProGuard is a free Java class file shrinker, optimizer, and obfuscator. It can detect and remove unused classes, fields, methods, and attributes. It can then optimize bytecode and remove unused instructions. Finally, it can rename the remaining classes, fields, and methods using short meaningless names. The resulting jars are smaller and harder to reverse-engineer."

We chose this application for the same reasons that we chose JReversePro. ProGuard is also bigger than JReversePro, with 351 classes and about 37 000 lines of code. The three functionalities rely on the same internal representation of Java programs that ProGuard builds from the jar files. Our hypothesis was that the slicing of any one of the three functionalities would result into the inclusion of the classes, methods, and attributes that are exclusive to that one functionality, along with the classes, methods and attributes required to build and navigate the internal representation model.

Much like JReversePro, ProGuard's command line interface was contained in a single class, ProGuard, and the function calls were dispatched in a single method, execute, that used a control parameter to dispatch to the proper function. This time though, the function call could not be used as a slicing criterion because the functions did not return any result. Instead, each function saved its result in the in-memory representation of the analyzed class files (the input to ProGuard). Accordingly, we had to inspect the code to find the place where the final result was produced (e.g. the shrunk version of the code), before it was written on the internal representation, and use *that* as a slicing criterion. We also had to comment the call to other functions found in conditional statements in the execute method because we did not want to record their side effects: for example, to extract the shrink functional aspect, we commented out the call to the optimize and obfuscate functions. We will discuss this issue in more detail in section 4.3, and see how this could be done systematically, but having done that manually, we were able to extract and validate the corresponding functional aspects. Table 1 shows the number of classes extracted by the slicer and the number of classes exclusive to each functional aspect:

|  | Sliced classes | Excl. classes |
| --- | --- | --- |

| Original program | 351 | 351 |
| --- | --- | --- |
| Shrinking | 165 | 30 |
| Optimizing | 232 | 109 |
| Obfuscating | 84 | 52 |

**Table 1. Functional Aspects size for ProGuard**

### 4.3. Issues

The selection of a slicing criterion is crucial to the quality of the extracted slice. This manifested itself in many ways. First, if we want to capture a functionality in its entirety, but only that functionality, we have to find the place in the program where the final result of the functionality—if such a thing exists—is produced. In our case studies, we encountered two reasonably well-behaved programs, both of which had main programs, and both of which dispatched their main functionalities in the main program, or a couple of calls further down the call graph. It was then possible to select the appropriate slicing criterion. However, with both JReversePro and ProGuard, we had to modify the source code by commenting out alternate branches of a conditional statement.

In the general case, we wouldn't be as lucky:
1. not all applications have a main program, or a single entry point. In other case studies, we had to write our own "main" program to create a suitable slicing criterion
2. not all functions that invoke a given functionality return a single and meaningful result. In the case of ProGuard, the three functionalities modified an object in memory that was internal to the running program, as opposed to returning a result. We refer to these as side-effectal functions.

One way to address side-effectal functions is to find the statements where the functions write those side effects, and slice on those statements. If a function has many side effects (e.g. modifying the values of several attributes), we can compute several slices, one per side effect, and merge them. We can also artificially create a single slicing criterion inside a bogus function that is called (through Aspect weaving) after each attribute modification. Either way, we need to develop some intimate knowledge of the internal structure of the program to do this properly, and we are exploring ways to make this process partially automated.

### 5. Discussion

Empirical studies on program slicing have shown that the slices generated from strongly cohesive programs tend to be fairly large since most program statements are geared towards the computation of one or a few results [20]. By contrast, the basic premise of aspect-oriented software development is that any program of some complexity will embody a number of interwoven functional and architectural aspects. For our case studies, we purposely selected applications that were likely to embody different functional aspects, and our results showed that, with a careful selection of the slicing criterion, we are able to precisely delineate the various functional aspects. There is no contradiction between the results of empirical studies on slicing and our results; it is simply a matter of scale: (even well-designed) large applications *will* embody *several* cohesive functionalities.

This raises the more general question of how program design quality impacts functional aspect separability. At a microscopic level, we saw how the slicer was vulnerable to control coupling[4], adversely affecting the separability of the functional aspects. In such cases, we may be able to enhance the slicing algorithm with finer analyses (e.g. constant propagation) to obviate the need for manually altering the program source code. At a same time, we saw how the slicer is resilient to low-cohesion: if a function performs many unrelated computations, the slicer will isolate those parts that are relevant to the slicing criterion. This enabled us, among other things, to slice multiply-defined functions between the various aspects.

A more intriguing question is how design patterns affect slicing, and by extension, aspect separability. At first glance, it appears that the user of design patterns could affect slicing and separability both ways. On the positive side, a number of behavioral patterns support behavioral evolution by enabling objects to acquire new behaviors over time. This is done by encapsulating the new behavior separately from the host class. A number of patterns come to mind, including: decorator, observer, strategy, and visitor. On the negative side, we feared that the design patterns would derail the slicer, for several reasons. One reason is the extensive use of polymorphism through subclassing or through the java **implements** relation between classes and interfaces. Another reason is the fact that design patterns often introduce additional levels of indirection in control flow. With the observer/observable, for example, it is hard to trace the execution flow, even with strongly typed listeners. A third reason is the inversion of control that we might find in some frameworks ("we will call you, don't call us"). Etc.

---

[4] Control coupling between two procedures corresponds to the case where procedure A controls the execution of procedure B through a control parameter that A passes to B. With JReversePro, we had two such control parameters: the first, `OP_CODE`, between the main program and the dispatch function, and the second, a boolean that was used to determine whether to disassemble or decompile.

Interestingly, the ProGuard software uses the visitor pattern extensively. Recall that the visitor pattern enables us to add a polymorphic behavior to a class hierarchy, without modifying the class hierarchy itself. It does so by encapsulating all the implementations of that behavior in one class—the visitor. This supports feature oriented programming whereby each feature is implemented as a visitor. ProGuard used the visitor patterns to 1) perform analysis on the source code by marking nodes appropriately and gathering some statistics and 2) modify the source code (e.g. shrink, obfuscate, etc.). Slicing worked well in this case, thanks to Indus's precise points-to analysis, which identified only the visitors that were initialized and used. The dependency analysis determined that some visitors (e.g. ClassFileShrinker) modified the output, i.e., the source code representation, and included them in the slice. Since those *modifier visitors* needed the information gathered from the *marker visitors*, those too were included in the slice. Finally, interestingly, for some functional aspects, several methods in the visitor interface were removed since they were not used and called anywhere in the extracted slices.

This led us to take a deeper look at the impact of design patterns on the separability of functional aspects in general, and the precision of slicing. To this end, we identified the core OO language constructs and design techniques used in design patterns (e.g. polymorphism), and studied their impact on slicing on general, and on Indus in particular. The results are beyond the scope of this paper.

## 6. Related Work

There has been quite a bit of work on aspect mining [5] for the past half a dozen years. The Aspect Mining Tool (AMT [9]) combines lexical matching with basic type analysis to look for code patterns that would result from an aspect weaving *à la* AspectJ, i.e. crosscutting concerns tangled with other concerns. Idem with Tonella & Ceccato, who use conceptual clustering algorithms to classify execution traces [20]. Feature location techniques such as SNIAFL [24] or Dynamic Feature Traces [7] try to match high level feature descriptions with low-level code elements. The result is usually a ranked list of software elements that are rarely self-contained, i.e. that do not contain all the required dependencies. A major disadvantage of these approaches is that they are heuristic in nature: we cannot be sure if a matched element is indeed part of an aspect, or that all of the required elements are, indeed, present (matched). We do not have this problem with our approach: the slicing extracts everything that is needed by a particular feature—and possibly more. One advantage of feature location techniques is that they highlight important or exclusive elements in a particular feature. Our diff & reporting tool (see sample output in Table 1) does some of that.

The Executable Concept Slices approach [4] retrieves executable code fragments that correspond to a particular concept or feature through program slicing. An important contribution of that work is the automatic selection of slicing criteria using *Hypothesis-Based Concept Assignment*. Our work complements this study by: 1) characterizing the extracted slices, 2) assessing the quality and relevance of extracted code elements. Program slicing has been used extensively in program understanding case studies (such as [3]). However, little is said in those studies about how to select the slicing criteria, or the various workarounds to deal with program slicing limitations.

A number of researchers have started to tackle *aspect refactoring* [15], of which aspect mining is only the first step: once we have identified the code fragments at a given program site that we think are part of an aspect, we need to, 1) repackage them as an aspect in the target aspect language, and 2) replace them by an introduction ('weaving') of that aspect in that program site. Notwithstanding intrinsic weaknesses in the existing aspect refactoring approaches [23], they all focus on crosscutting concerns using AspectJ-like aspects and join points model.

This raises the related question of whether *object-oriented* refactoring methods and tools can be used to re-modularize a legacy application to separate functional aspects. First, the boundaries of that functional aspect have to be delineated, and that is what program slicing does for us. The interesting question is whether there exists an *object-oriented* refactoring of the source program that would isolate the functional aspect. Some early critics of aspect-oriented programming argued—but not formally proved—that the answer was affirmative, and that aspect oriented programming was not strictly needed. The visitor pattern is an example of an *object-oriented pattern* that separates functional aspects, and is not the only one. However, there is no guarantee that such an object-oriented pattern can always be found, or that the existing object-oriented refactoring methods have the required granularity to handle them: program slicing can slice individual instructions within methods!

## 7. Conclusion

In this paper, we studied the problem of extracting functional aspects from legacy OO code for the purpose of 1) easing program understanding and maintenance and 2) possibly repackaging these aspects for separate reuse using a variety of OO and non-OO techniques (e.g. HyperJ). We used the Subject/HyperJ conceptual model and the 'refers-to' relationships to characterize functional aspects and developed our own toolset around the Indus slicer [12] to extract aspects using program slicing.

Our case studies showed that program slicing can, indeed, help extract functional aspects that are more complete than current feature location techniques and more cohesive than known aspect mining approaches. Accessorily, we also assessed the impact of program design quality over program slicing and proposed some workarounds when needed. Since developers still need do to some work beforehand to prepare the program for slicing, we are currently developing additional tools to help automate the process. We are also conducting additional case studies on different kinds of applications to see if there are any qualitative differences in the supported functionalities.